\documentclass[	DIV=calc,%
							paper=a4,%
							fontsize=12pt,%
							twocolumn]{scrartcl}	 					
\usepackage{natbib}

\usepackage[english]{babel}										
\usepackage[protrusion=true,expansion=true]{microtype}				
\usepackage{amsmath,amsfonts,amsthm}					
\usepackage[pdftex]{graphicx}									
\usepackage[svgnames]{xcolor}									
\usepackage[hang, small,labelfont=bf,up,textfont=it,up]{caption}	
\usepackage{epstopdf}												
\usepackage{subfig}													
\usepackage{booktabs}												
\usepackage{fix-cm}													
\usepackage{hyperref}
\hypersetup{
  colorlinks=true,
  linkcolor=DarkGreen,
  citecolor=DarkGreen,
  filecolor=DarkGreen,
  urlcolor=DarkGreen
}

\usepackage{sectsty}													
\allsectionsfont{
	\usefont{OT1}{phv}{b}{n}
	}

\sectionfont{
	\usefont{OT1}{phv}{b}{n}
	}

\usepackage{fancyhdr}												
\usepackage{lastpage}	
\usepackage{listings}
\usepackage{spverbatim}

\newcommand{\apjs}{Astrophysical Journal Supplement}
\newcommand{\apjl}{Astrophysical Journal Letters}

\newcommand{\aap}{Astronomy and Astrophysics}

\usepackage{lettrine}
\newcommand{\initial}[1]{%
     \lettrine[lines=3,lhang=0.3,nindent=0em]{
     				\color{DarkGreen}
     				{\textsf{#1}}}{}}

\usepackage{titling}															

\newcommand{\HorRule}{\color{DarkGreen}
									  	\rule{\linewidth}{5pt}%
										}
\pretitle{\vspace{-30pt} \begin{flushleft} \HorRule 
				\fontsize{40}{40} \usefont{OT1}{phv}{b}{n} \color{DarkGreen} \selectfont 
				}
\title{{\Huge  Can AI Understand Our Universe?} \\ {Test of Fine-Tuning GPT by Astrophysical Data}}				
\posttitle{\par\end{flushleft}\vskip 0.5em}

\preauthor{\begin{flushleft}
					\large \lineskip 0.5em \usefont{OT1}{phv}{b}{n} \color{Black}}
\author{Yu Wang $^{1,2,3}$, Shu-Rui Zhang $^{1,2,4}$, Aidin Momtaz $^{5}$,  Rahim Moradi $^{3, 6}$, Fatemeh Rastegarnia$^{1,2,7}$, Narek Sahakyan  $^{1,2,8}$, Soroush Shakeri $^{1,2,9}$ and Liang Li $^{1,2,3}$.}
\postauthor{\footnotesize \usefont{OT1}{phv}{m}{sl} \color{Black} \\\vskip 0.5em 1. International Center for Relativistic Astrophysics (ICRA), Roma I-00185, Italy \\ 2. International Center for Relativistic Astrophysics Network (ICRANet), Pescara I-65122 , Italy \\ 3. National Institute for Astrophysics (INAF) -- Osservatorio Astronomico d'Abruzzo, Teramo I-64100, Italy \\ 4. Department of Astronomy, University of Science and Technology of China, Hefei 230026,  China \\ 5. Department of Physics, University of Padova, Padova I-35122, Italy \\ 
6. Institute of High Energy Physics, Chinese Academy of Sciences, Beijing 100049, China\\7. Dip. di Fisica e Scienze della Terra, Universit\`a degli Studi di Ferrara, Ferrara I--44122, Italy\\ 8. ICRANet-Armenia, Marshall Baghramian Avenue 24a, Yerevan 0019, Armenia \\ 9. Department of Physics, Isfahan University of Technology, Isfahan 84156-83111, Iran   
\\ Emails: wang@inaf.it, zhangsr@mail.ustc.edu.cn, rmoradi@ihep.ac.cn, liang.li@icranet.org 
\par\end{flushleft}\HorRule}

\date{}																				

\begin{document}
\maketitle
\thispagestyle{fancy} 
\initial{C}\textbf{hatGPT has been the most talked-about concept in recent months, captivating both professionals and the general public alike, and has sparked discussions about the changes that artificial intelligence (AI) will bring to the world. As physicists and astrophysicists, we are curious about if scientific data can be correctly analyzed by large language models (LLMs) and yield accurate physics. In this article, we fine-tune the generative pre-trained transformer (GPT) model by the astronomical data from the observations of galaxies, quasars, stars, gamma-ray bursts (GRBs), and the simulations of black holes (BHs), the fine-tuned model demonstrates its capability to classify astrophysical phenomena, distinguish between two types of GRBs, deduce the redshift of quasars, and estimate BH parameters. We regard this as a successful test, marking the LLM's proven efficacy in scientific research. With the ever-growing volume of multidisciplinary data and the advancement of AI technology, we look forward to the emergence of a more fundamental and comprehensive understanding of our universe.\newline\indent This article also shares some interesting thoughts on data collection and AI design. Using the approach of understanding the universe—looking outward at data and inward for fundamental building blocks—as a guideline, we propose a method of series expansion for AI, suggesting ways to train and control AI that is smarter than humans\footnote{Code will be available in the next revision.}.}

\section*{Movitation}

Most researchers in the field of science first encountered the concept of deep learning around the year 2012 \citep{NIPS2012_c399862d}. At that time, the renaissance  of neural networks just started, and the community was primarily focused on improving the accuracy of object recognition in images. Nearly no one anticipated the appearance of the attention mechanism that would make convolutional neural networks (CNNs) more versatile and it brought the birth of a universal structure known as the transformer in a decade \citep{NIPS2017_3f5ee243}. The stacking of transformers led to a qualitative change triggered by a quantitative increase: with the explosive growth in the number of parameters, machines seemed to genuinely possess intelligence \citep{2023arXiv230312712B}. The pace of development over these ten years is likely beyond what even the most optimistic individuals at the time could have imagined. 

However, at the same time, we clearly recognize that AI is still in its infancy. The first cry of this infant has shocked us, even frightened us, as humans have created a potential intelligent entity capable of replacing humans. What expectations should we have for this infant, how should we educate it, and what impact will it have on our world are immense questions that challenge to provide a definitive answer. Yet, every person interested in machine learning can't help but ponder them. As physicists and astronomers, we inevitably ask: Can AI understand our Universe?

\subsubsection*{Paths of Understanding Our World}

Physics has been developed over for 400 years, especially since the birth of relativity and quantum mechanics in the last century, the ultimate goal of physicists has generally been to seek the fundamental principles governing everything. 
\begin{equation}
    \mathcal{L} = \mathcal{L}_{SM} + \mathcal{L}_S + \mathcal{L}_\Lambda + \mathcal{L}_N + \mathcal{L}_\varphi - V_{RH} ~+ ~?
\label{eq:lagrangian}
\end{equation}
Equation \ref{eq:lagrangian} is the Lagrangian of known physics \citep{2005PhLB..609..117D}, with the first term $\mathcal{L}_{SM}$ representing the Standard Model. It encompasses strong and electroweak interactions, CP violation, gravity, supersymmetric particles, Higgs field, and their interrelations, among others. These are aspects that we understand relatively well and can model. The later terms signify those elements that we know exist but whose correct forms are yet to be confirmed. These include dark matter $\mathcal{L}_S$, dark energy $\mathcal{L}_\Lambda$, neutrino masses $\mathcal{L}_N$, the birth of the universe $\mathcal{L}_\varphi$, and how particles were formed post-creation $V_{RH}$. Of course, this equation is still incomplete, there might be more terms that we don't even know exist.

Traditionally, the development of physical theories described in equation \ref{eq:lagrangian} by humans has followed two main paths. The first path is based on observation: collecting data, summarizing experience from the data, and then abstracting it into theory. The heliocentric theory of Galileo and Darwin's theory of evolution are both grounded in observational data. The other path is based on philosophical foundations or a broader framework of thought to guide the construction of theories. Notable examples include Aristotle's view of physics and Einstein's theory of relativity.

How have modern physics and astrophysics developed? Firstly, most scientists agree that data and philosophy are complementary. The general consensus is that data alone is insufficient, human understanding is needed to interpret the data and human logical and dialectical abilities can abstract theories from data. However, current scientific development relies more on data than before, whether in physics, biology, social sciences, or other fields. This reliance on data is a characteristic of our era of rapid technological advancement. Observational developments are incredibly fast, as evidenced by the Lagrangian of equation \ref{eq:lagrangian}, where the terms of unknown form were discovered through observation, while the theory is still being constructed. 

\subsubsection*{Paths Integrating AI}

What can AI contribute to the paths of understanding?

It is widely believed that AI has an advantage over traditional data analysis methods. AI can analyze big data in a short time that traditional methods can't handle, and it can identify information in detail that is difficult for traditional methods to recognize \citep{najafabadi2015deep,jan2019deep}. If observational data is insufficient, AI can generate simulated data. For those simulation tasks that currently require supercomputers for days or even months, a trained AI can complete them within a few seconds \citep{2023arXiv230204062L}. For experiments with ethical constraints or potential harm to humans, such as drug trials, AI can perform molecular screening and docking, as well as pharmacokinetic simulations \citep{dara2022machine}.

The establishment of theories firstly requires logic. After the rise of LLMs, it is widely believed that AI has acquired the capability of logical reasoning. The similarities and differences between AI logic and human logic are currently a topic of discussion. Secondly, in terms of mathematical ability, current AI has found applications in fields such as automated theorem proving, number theory, and graph theory, and many mathematicians are optimistic about AI's future abilities in mathematics \citep{2021Natur.600...70D,2024Natur.625..468R}. The creation of theories often requires innovative ideas. Some believe that AI could display novel scientific intuition and creativity in forming hypotheses, akin to human rationalist approaches. However, this is debatable at the current stage, with one challenge being the difficulty in quantitatively assessing creative thinking. From a practical standpoint, AI is capable of participating in brainstorming and generating ideas, offering a multitude of perspectives or suggestions on a given topic.

The integration of AI's capabilities in handling vast knowledge and data across multiple areas, combined with its logical reasoning and potential for creative thinking, heralds a new era of scientific research.  It represents a shift from specialized to integrated knowledge, promising a more interconnected and fundamental understanding of the natural world.

\subsubsection*{Large Language Model}

LLMs represent a significant advancement of AI in the recent a few years. The term ``large" is twofold: Firstly, it reflects the model's training on an expansive, diverse dataset that encompasses a vast spectrum of human knowledge. This dataset includes a variety of text sources such as books, articles and websites covering an extensive array of fields. Secondly, ``large" pertains to the architecture of the model, specifically its number of parameters. Models like GPTs \citep{radford2018improving,radford2019language,brown2020language} boast billions or even trillions of parameters, a crucial feature for their ability to memorize information and facilitate logical information processing. The ``language" signifies the training and output of the model are presented by human languages.

LLMs seem to possess the potential as an AI analyzing data and theoretical reasoning by its following characteristics:

\begin{enumerate}
    \item Language is not only a tool for communication but also a medium for thinking. Thinking relies on knowledge, which is essentially filtered data.
    \item Language itself contains logic, and is capable of contextual logical reasoning and judgment. Models trained with data represented in language can inherit the intrinsic logical characteristics of language.
    \item The Transformer architecture used by the LLMs is a general-purpose network architecture capable of processing various types of data and tasks.
    \item LLMs can use plugins or write and execute code to perform precise actions.
\end{enumerate}

In the following sections, we first introduce the techniques for fine-tuning the LLM of GPT model. Then we fine-tune the GPT model by astronomical data and test the capacity of the fine-tuned GPT to complete multiple astrophysical tasks, this is a preliminary test of whether large language models have the potential to understand the universe. It should be noted that here we fine-tune a general model for multiple tasks, rather than designing specific models for specific tasks. We will also discuss the interdependence and promotion between AI and big scientific data acquired by future large scientific facilities. Finally, we will make a brief summary of the article.

\section*{Fine-Tuning Procedure}

Our article tests the performance of GPT for a few selected types of astronomical data. We believe there are more astrophysicists who want to test their own data or explore deeper into the capabilities of GPT. Therefore, we try to demonstrate the technical process of fine-tuning, and we also open source all the codes and data involved in the article, wishing to present an informative article for other peers.

\subsubsection*{Pre-traing and Fine-tuning}
Training ChatGPT, or any LLM like it, involves a two-step process\footnote{The reinforcement learning through human feedback is neglected in this article because we don't perform this step for the astronomical data.}: pre-training and fine-tuning \citep{radford2018improving}. 

During the pre-training phase, the model learns to predict the next word in a sentence based on the words that came before it. This is achieved by training the model on the collected large dataset using a technique called Masked Language Modeling. This dataset typically consists of a diverse range of sources, including books, articles, and websites. The goal is to create a representative sample of human language that can be used to teach the model grammar, facts, and reasoning abilities. The model learns to understand grammar, syntax, and context from this process, but it may not be fine-tuned to specific tasks or guidelines.

After the pre-training, the model needs to be fine-tuned to perform specific tasks, like answering questions or generating text. To do this, a smaller, more specific dataset is created, usually by gathering examples of the desired behavior. The dataset often consists of input-output pairs, where the input is a prompt, and the output is a desired response. This training step is supervised by humans.

\subsubsection*{Tokenization}

Before preparing the data, we first introduce the data presentation of LLM, the Tokenization \citep{mielke2021between}.

Tokenization is the process of breaking down text into smaller units, called tokens. Tokens can be words, subwords, or individual characters, depending on the tokenization strategy employed. The purpose of tokenization is to convert raw text into a format that can be easily processed by the LLM, assigning numbers to each token and using those numbers to encode text.

GPT adopts the subword-level tokenization technique, Byte Pair Encoding (BPE) \citep{gage1994new}, to alleviate the problem that Word-level tokenization can lead to a large number of out-of-vocabulary (OOV) words. BPE works by iteratively merging the most frequent character pairs in the training data until a desired vocabulary size is reached. This results in a vocabulary that contains a mix of common words, subwords, and individual characters. BPE brings the advantage of allowing GPT to work with a smaller vocabulary, leading to reduced memory requirements and computational costs. BPE can also handle different languages, dialects, and linguistic variations, making it suitable for multilingual models like GPT.

\subsubsection*{OpenAI API}
OpenAI provides the API\footnote{API reference: \url{https://platform.openai.com/docs/api-reference/introduction}} to access and to fine-tune the pre-trained models of Babbage, Davinci, gpt-3.5-turbo and gpt-4 (exprimental)\footnote{These are the models provided at the time of writing this article. OpenAI will update the models, please check the official website for the new list.}. The model size and the computational requirements increase from Babbage to gpt-4, as well as the price\footnote{Fine-tuning price: \url{https://openai.com/pricing}}. 

For users who want to test the LLM models for their specific purpose, the easiest way is to implement the OpenAI API. The users only need to care about the data preparation for fine-tuning purposes and to upload the formatted data to the OpenAI server, OpenAI will do the training work automatically. 

In general, our fine-tuning procedure follows the OpenAI documentation\footnote{Fine-tuning documentation: \url{https://platform.openai.com/docs/guides/fine-tuning}}, the steps are:
\begin{enumerate}
\item Prepare the training and test data, just like for a general neural network training task, with both features and labels. For instance, each feature is a list containing 1000 values, and each label is a single value. To save token usage, some transformations can be applied to the data. We will show in the following section of Application the example that each number is multiplied by 1000 and then rounded. The specific transformation method should be formulated according to the rules of tokenization and the actual problem. Some useful tokenization rules include: three digits count as one token, "0." counts as two tokens, a comma is one token, etc.

\item Transform the data into OpenAI's data format. Each column of data consists of three parts. The first is \texttt{system}, which describes the problem; next is \texttt{user}, which contains the content of our feature; and lastly \texttt{assistant}, which contains the  label. All data are stored in the JSONL format. An example of each column of data is as follows:
\begin{spverbatim}
   {"messages": [{"role": "system", "content": "from SDSS optical spectrum, infer redshift of one value:"}, {"role": "user", "content": "1000 85 69 47 47 43 14 22 48 63 28 46 48 55 54 65 80 61 61 0 58 43 52 40 68 66 65 54 80 77 76 73 96 67 77 95 84 105 107 99 110 115 124 108 127 106 125 134 118 101 151 111 109 137 122 150 205 210 225 193 221 240 239 227 256 276 288 311 318 333 324 310 316 341 334 350 332 362 333 331 304 327 373 386 388 388 369 383 362 388 403 408 410 337 409 395 442 345 311 404"}, {"role": "assistant", "content": "0.597"}]} 
\end{spverbatim}

\item Upload the data to OpenAI and start training by setting up a fine-tuning task, with code references available on the official website and the examples we provide. Key parameters include \texttt{api\_key}: your personal or institution's API, which can be obtained from account information\footnote{API keys: \url{https://platform.openai.com/api-keys}}; \texttt{model}: the model name, which can be an initial model provided by OpenAI such as \texttt{gpt-3.5-turbo}, or the name of an already trained model, for example, \texttt{ft:gpt-3.5-turbo:personal::9oOG13x}; \texttt{n\_epochs}: the number of epoches the data is trained, different from the hundreds of epochs for ordinary tranings, fine-tuning GPT usually taking 1 or 2. Training will generate a job id, and you can query the progress and other information of the training based on the job id. If OpenAI's servers are busy, you may need to wait for several tens of minutes before starting training. To train 100 pieces of data, approximately less than 3 minutes are needed, the exact time depends on the data and server conditions.

\item Testing and inference can be implemented through the API or can be done in the web-based playground\footnote{Playground: \url{https://platform.openai.com/playground/chat}}.
\end{enumerate}





\section*{Applications}

\subsection*{Classification of Quasar, Galaxy, Star and BAL}\label{subsec:classification}

\begin{figure*}
\centering
\includegraphics[width=1.0\hsize]{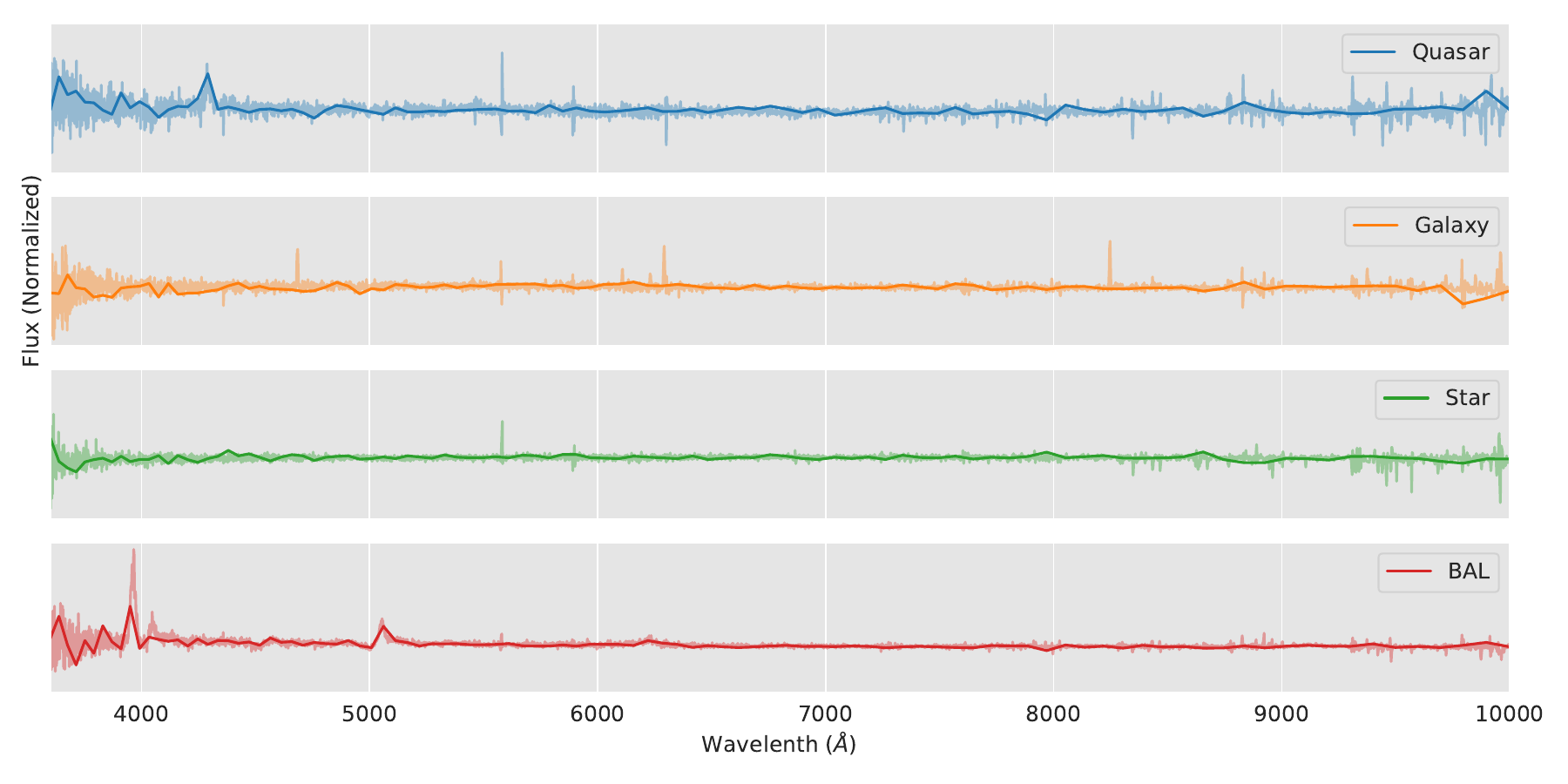}
\caption{Example of quasar, galaxy, star, and BAL spectra. The lighter colors depict the original high-resolution spectral data, while the darker colors represent the downsampled version, consisting of 100 data points for each spectrum.}\label{fig:spectra}
\end{figure*}

\begin{figure}
\centering
\includegraphics[width=1.0\hsize]{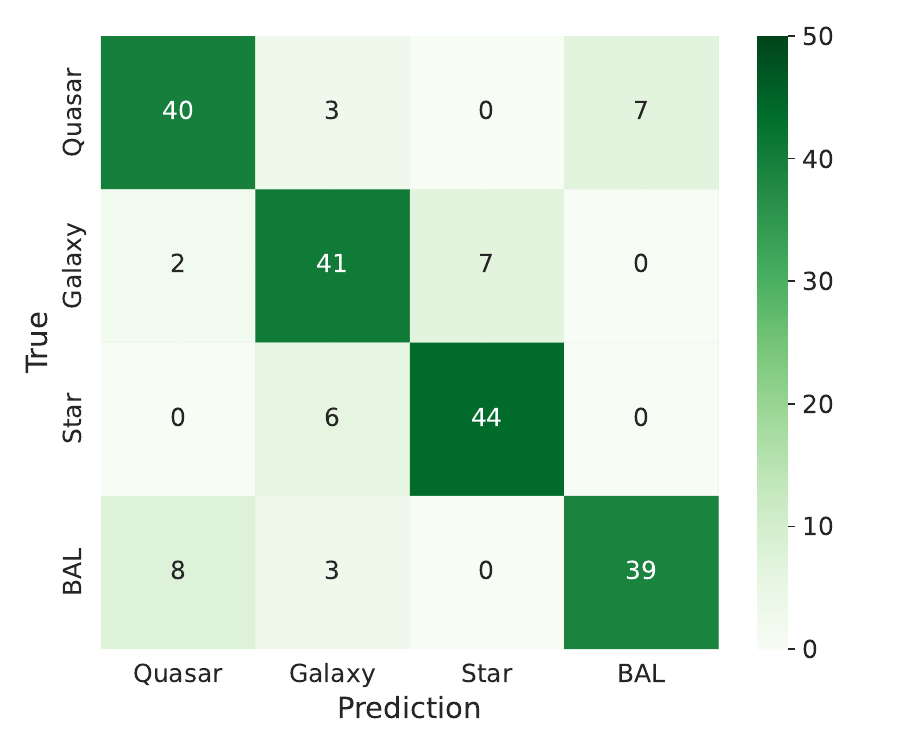}
\caption{Confusion matrix of four SDSS classes. Indicating, for e.g., 40 quasars are correctly predicted as quasar, 3 quasars are wrongly predicted as galaxy, and 7 quasars are wrongly predicted as BAL.}\label{fig:confusion-matrix}
\end{figure}

 The Sloan Digital Sky Survey (SDSS) is a major imaging and spectral survey of our Galaxy and Universe \citep{2000AJ....120.1579Y}. SDSS collects data to map the Universe, studying a vast array of cosmic entities including quasars, galaxies, stars, and other celestial objects. The survey has gone through four phases and is currently undergoing the fifth phase, each with specific goals. The primary objectives of SDSS-I (2000-2005) were to map the large-scale structure of the Universe through a redshift survey of galaxies and to detect and characterize quasars  \citep{2006ApJS..162...38A}. SDSS-II (2005-2008) mapped the structure and kinematics of the Milky Way using 240000 stellar spectra, and monitored a 300 square degree region to detect and measure supernovae, contributing to the understanding of dark energy \citep{2009ApJS..182..543A}. SDSS-III (2008-2014) collected detailed spectra in the infrared for 100000 Milky Way stars to study the chemical and kinematical history of our galaxy, measured the redshifts of 1.5 million luminous galaxies and 160000 quasars to study baryon acoustic oscillations and the cosmic distance scale, and searched for exoplanets by obtaining radial velocity measurements of 8500 stars \citep{2015ApJS..219...12A}. SDSS-IV (2014-2020) extended the quasar and galaxy observation focusing on cosmic evolution over the last six billion years and the behavior of dark energy, also extended the study of the Milky Way into the Southern Hemisphere and added a time-domain component, and it used spatially resolved spectroscopy to study the internal kinematics of a large sample of 10000 nearby galaxies \citep{2022ApJS..259...35A}. SDSS-V (2020-2025) aims to provide a comprehensive census of black holes, from the smallest to the most massive, by observing them through various methods, including reverberation mapping, to study the assembly history and evolution of the Milky Way using stellar spectra, and to provide a detailed understanding of the physical processes that drive galaxy evolution by studying galaxies in the nearby universe \citep{2023ApJS..267...44A}. 

Over time, SDSS has significantly expanded in scale and scope, each phase contributing to an ever-growing dataset. Data releases (DR) of SDSS are periodic updates where collected astronomical data is made publicly available. Each DR is a cumulative collection that includes all past data plus new observations. They contain images, spectroscopic data, and catalogs, aiding in various astronomical studies. Key components often include spectra and images of stars, galaxies, and quasars, along with derived measurements. The latest data release is DR18, which is part of SDSS-V \citep{2023ApJS..267...44A}. Our samples are constructed based on DR16, which published the observations of SDSS-IV and accumulated the observations of previous phases. 

We start with the dataset prepared in a previous work \citep{2022MNRAS.511.4490R}, this dataset contains all the spectra with visual-inspected redshift from DR16. The selected sources are catalogued to quasars, galaxies, stars, and broad absorption line (BAL) quasars. The numbers of these four classes are 396831, 30666, 3406 and 39501, respectively. The spectra of each class are combined and stored in a single file and have been uploaded to Kaggle\footnote{\url{https://www.kaggle.com/datasets/ywangscience/sdss-iii-iv}}. 

To limit the cost of fine-tuning, we extracted 500 samples from each class, totaling 2000 samples. We then re-sampled our spectral resolution within a logarithmic domain of 360 nm to 1000 nm, settling on 100 equidistant data points, exampled in figure \ref{fig:spectra}. We further optimized the number of tokens\footnote{Tool to estimate token number: \url{https://platform.openai.com/tokenizer}} by first  we normalized all spectral values to lie within the 0-1 range. Subsequently, we retained only the initial three fractional digits. As a result, the processed spectra manifested as sequences of 100 3-digit numerals, delineated by spaces, for instance, ``178 351 396 ... 168 101 456''.

From the 2000 samples, we allocated 1800 for training and 200 for testing. We fine-tuned the GPT-3.5 turbo model using the training set over two epochs. Post-training, our test data inference yielded an overall accuracy of 82\%, with quasars at 80\%, galaxies at 82\%, stars at 88\%, and BALs at 78\%. This implies that with only 1800 samples for fine-tuning, we achieved a commendable accuracy, demonstrating the GPT model's efficacy in classifying SDSS spectra. The confusion matrix, depicted in figure \ref{fig:confusion-matrix}, reveals that the model tends to confuse quasars with BALs and galaxies with stars. However, distinguishing between quasar and BAL, as well as galaxy and star are relatively more accurate. 

It's noteworthy that in order to constrain the cost, we utilized less than 5\% of the total spectra and reduced the sampling rate by an order of magnitude compared to the observations. We anticipate enhanced accuracy if the entire original spectral dataset are used to fine-tune the GPT model.

\subsection*{Inferring Redshift of Quasars}\label{subsec:redshift}
\begin{figure}[!t]
\centering
\includegraphics[width=1.0\hsize]{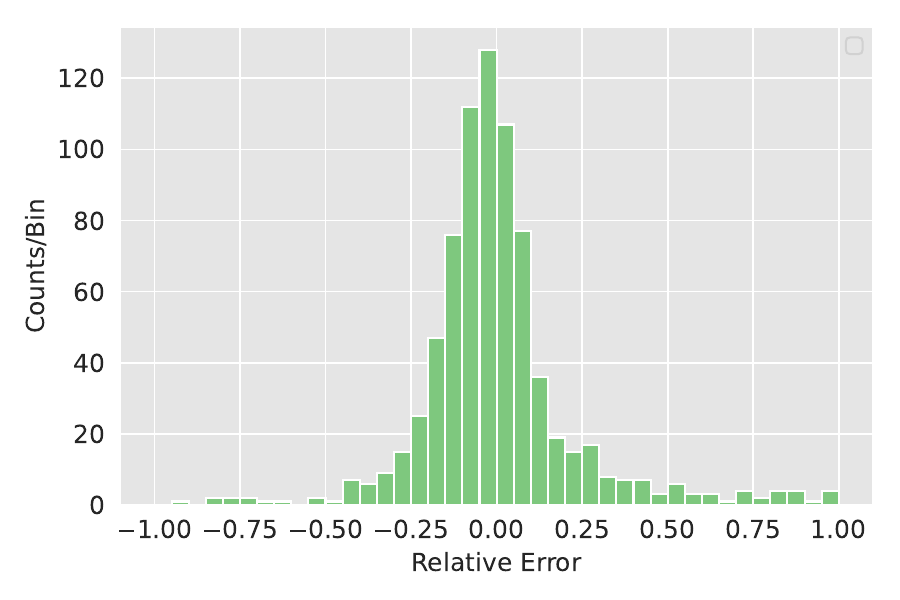}
\includegraphics[width=1.0\hsize]{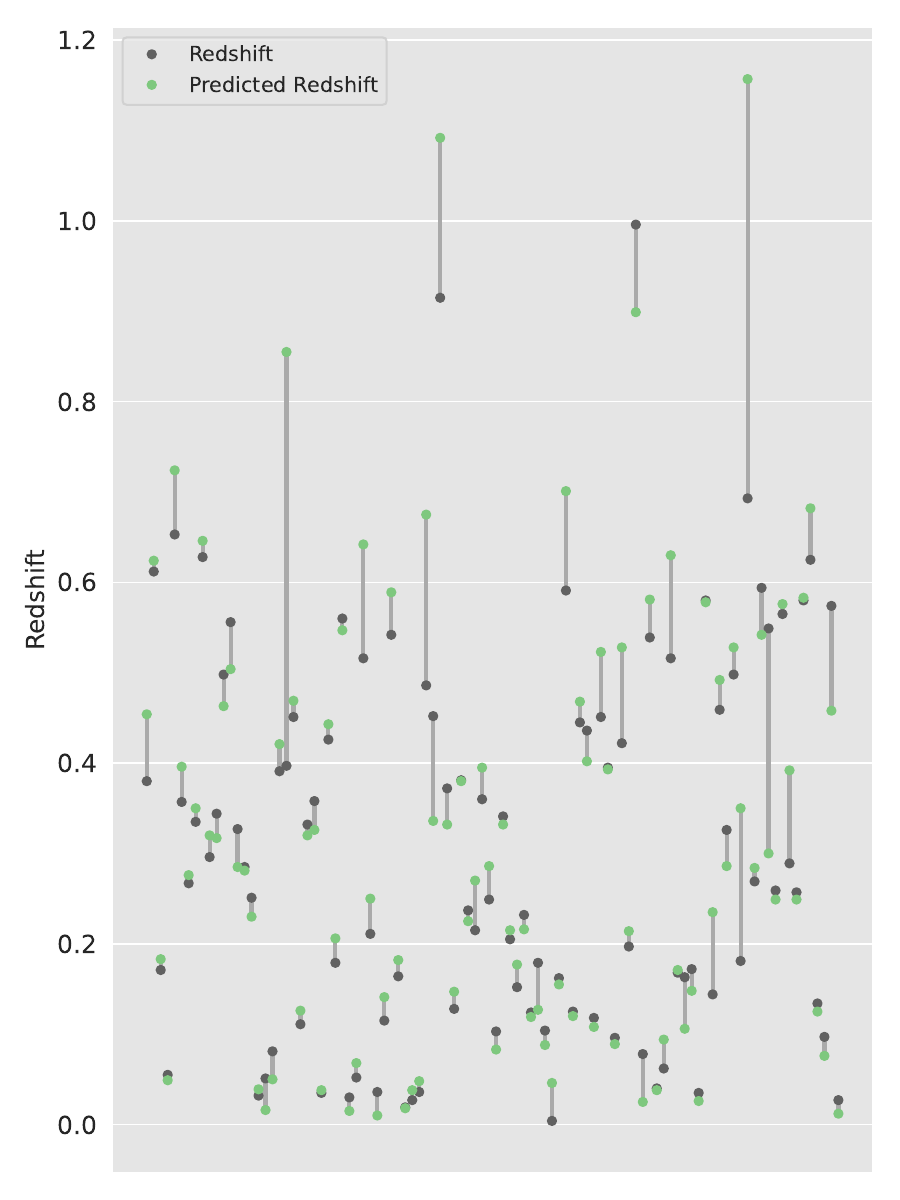}
\caption{Top: histogram of relative error of the predicted redshift. Bottom: example of redshift prediction of 100 random quasars.}\label{fig:redshift-examples}
\end{figure}

Quasars, powered by accretion of material onto supermassive black holes, serve as vital probes into the universe's history \citep{2006NewAR..50..665F}. Determining the redshift of quasars refines our measurements of key cosmological parameters, enhancing our grasp of the universe's expansion rate and dark energy properties. Quasars brightness and distribution make them excellent markers for mapping the universe's large-scale structure. Furthermore, the spectra of quasars provide insights into the interstellar and intergalactic medium, revealing details about gas clouds' composition and density. 

Especially, high-redshift quasars offer a glimpse into the universe's conditions during its early stages, allowing astronomers to investigate the first billion years post-Big Bang. Observations of early galaxies hosting supermassive black holes, formed surprisingly soon after the Big Bang, challenge our understanding of black hole growth rates and suggest rapid early star formation \citep{2024Natur.627..281A}. These distant quasars have also been instrumental in pinpointing the epoch of reionization, marking the universe's transition from neutrality to ionization \citep{2022NatAs...6..850J}. Their distribution provides insights into the universe's primordial large-scale structure \citep{2018A&A...612A..31P}. 

In this section, we utilize spectra from the SDSS, analogous to the previous section, but focus solely on quasar-classified spectra. Guided by the principle of assessing the efficacy of fine-tuning the GPT model for a specific astrophysical query within a constrained budget, rather than pursuing extreme accuracy, we sampled 4000 quasar spectra with redshifts ranging from 0 to 1 out of nearly 400k available. Of these, 3,200 were designated for training, while the remaining 800 were set aside for testing, only one epoch of training was performed. The fine-tuned model demonstrated a high accuracy, with a median relative accuracy rate of 90.66\%. Figure \ref{fig:redshift-examples} illustrates the  histogram of relative error rates and a subset of the test results. The fine-tuned GPT is capable of not only classification tasks as shown in the previous section but also effectively executing regression tasks as predicting a float value of the redshift.

For a more precise redshift estimation, a higher sampling rate may be essential. As observed from figure \ref{fig:spectra}, the current sampling of 100 data points omits numerous emission lines, which are fundamental to direct redshift calculations based on line shifts. Thus, by augmenting the sampling rate and sample size, we believe that the GPT model's redshift estimation can be further refined. It is worth noting that our sample is confined to the redshift range of 0 to 1. Hypothetically, spectra with higher redshifts, given adequate samples, would yield similar outcomes. Quasars with redshifts exceeding 6 are rare, necessitating the incorporation of simulated data and other specialized methodologies to enhance training. Such intricacies warrant further deliberation and experimentation, which are beyond the scope of this paper.

\subsection*{Short or Long Gamma-ray Burst}\label{subsec:grb}

\begin{figure*}
\centering
\includegraphics[width=1.0\hsize]{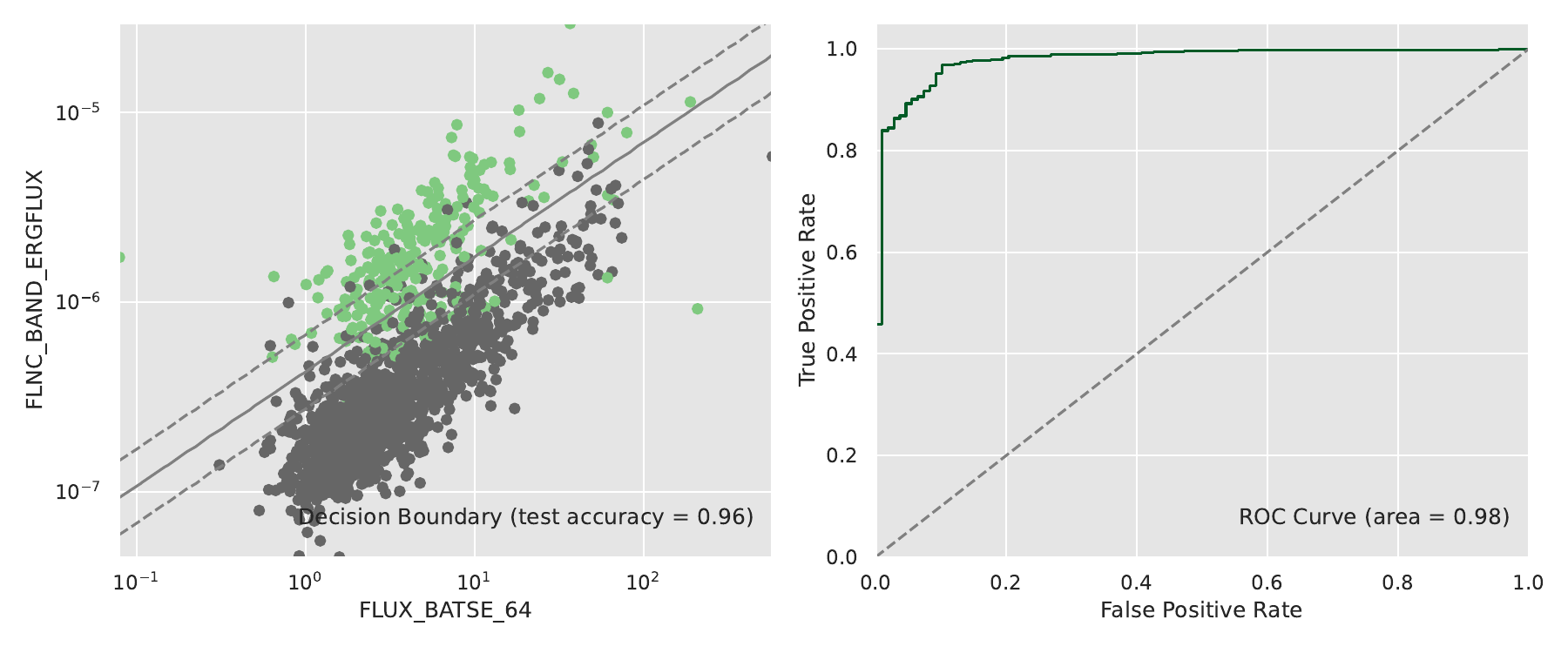}
\caption{Example of the feature selection procedure by SVM. Left: two parameters ('FLUX\_BATSE\_64' and 'FLNC\_BAND\_ERGFLUX') from the catalog are selected to compute the decision boundary and then to compute the accuracy of classification. Right: the corresponding ROC curve.}\label{fig:parameters}
\end{figure*}

Gamma-ray bursts (GRBs) can be broadly categorized into two types: short bursts and long bursts. This classification is traditionally determined by the duration parameter \(T_{90}\), where \(T_{90}\) is the time duration in which 90\% of the burst's total emission is observed. Specifically, bursts with \(T_{90} < 2\) seconds are considered short bursts, whereas those with \(T_{90} > 2\) seconds are designated as long bursts \citep{1993ApJ...413L.101K}.

It's notable, however, that the duration of both short and long GRBs conforms to a normal distribution. As a result, these two distributions overlap at certain points. This means that using a simple 2-second threshold for classification can misclassify some bursts \citep{2015NatCo...6.7323Y}, especially those located on the tails of the distributions. Consequently, there's a potential risk of incorrectly categorizing a GRB that may have a duration close to this threshold. Our primary objective in this study is to explore alternative observational metrics that can enhance the accuracy of classifying short and long GRBs, especially from the spectral information, not the time-related information. By examining various observational parameters comprehensively, we aim to rectify the potential misjudgments that arise from solely relying on the \(T_{90}\) criteria.

The dataset\footnote{Updated to Oct. 11, 2023.} employed for our research is derived from two primary catalogs: the Fermi GBM Trigger Catalog\footnote{\url{https://heasarc.gsfc.nasa.gov/W3Browse/fermi/fermigtrig.html}} and the GBM Burst Catalog \footnote{\url{https://heasarc.gsfc.nasa.gov/W3Browse/fermi/fermigbrst.html}} \citep{2016ApJS..223...28N}. The Catalogs are extensive tables listing observational parameters of 14 GBM detectors, consisting of 12 NaI and 2 BGO units. All GBM triggers are featured in the Trigger Catalog, but only those identified as bursts make it to the Burst Catalog. The Trigger Catalog has $11$ parameters that include mainly the time and location information, and the burst catalog of parameters $100$ contains the primary output of the duration metrics \(T_{50}\) and \(T_{90}\), and the parameters fitted from multiple spectral models, including power law (PL), the cutoff power law (CPL), the band and the smoothly broken power law (SBPL).

The Trigger Catalog contains a total of 9547 samples, each characterized by 29 distinct parameters. In contrast, the Burst Catalog is comprised of 3625 samples, detailed by an extensive 306 parameters. By intersecting these two catalogs and discarding extraneous parameters such as sample versions, the resultant dataset consists of 3485 GRBs described by 302 parameters. If we classify these GRBs solely based on the $T_{90}$ criterion, our dataset includes 598 short-duration bursts and 3027 long-duration bursts. It's pertinent to highlight that, due to observational constraints or fitting methodologies, not every GRB furnishes a complete set of parameter values.

Our analytical framework commences with ``Feature Selection'', wherein we discern a subset of pivotal parameters from the extensive list of  over 300 parameters. This stage aims to: 1) Extract salient features conducive to formulating the physical model of GRBs. 2) Enhance the accuracy and efficacy, especially by minimizing the incorporation of spurious or noisy features to amplify the predictive power and generalizability of neural networks. 3) Reduce the token amount, lowering the cost of fine-tuning.

For GRB classification via the T90 parameter, one would conventionally plot a histogram of T90 values and subsequently fit the histogram with two normal distributions. The intersection of these two distributions occurs just near 2~s, hence adopting 2~s as the demarcation between long and short bursts.  Our method, taking cues from empirical relations such as the Amati, Ghirlanda and Liang-Zhang relations \cite{2002A&A...390...81A, 2004ApJ...616..331G, 2005ApJ...633..611L}, we endeavor to discern whether any two parameters from the catalog can proficiently bifurcate long and short bursts. For instance, we chose 'FLUX\_BATSE\_64' and 'FLNC\_BAND\_ERGFLUX', which correspond to the peak flux in the BATSE standard 50-300 keV energy band (64 ms timescale) and the average energy flux for a Band function law fit an integrated spectrum over the burst's duration, respectively. As shown in the figure \ref{fig:parameters}, visually the long and short bursts are clustered into two groups, with some overlap. In a subsequent analytical step, we first apportion our dataset into training (75\%) and testing sets (25\%), then utilize the Support Vector Machine (SVM) methodology \citep{cortes1995support} to derive the decision boundary based on our training set, and subsequently evaluate its classification accuracy on the testing set. The decision boundary is evidenced by the solid line in figure \ref{fig:parameters}, and a receiver operating characteristic (ROC) curve is also shown there. The results show that these two parameters, 'FLUX\_BATSE\_64' and 'FLNC\_BAND\_ERGFLUX', achieve an overall classification accuracy of 95.68\%, with a short-duration burst accuracy of 87.04\% and a long-duration burst accuracy of 97.43\%.

Given the plethora of 91204 parameter combinations emergent from 302 parameters, we repeated the SVM training and inference process 91204 times, determining the classification accuracy for each combination. We then selected the combinations where both short and long burst accuracies are over 80\%. The top 15 most frequently appearing parameters in these optimal combinations are the outcome of our feature selection process, together with 10 parameters from the fitting of Comptonized model will be the final parameters used for the model fine-tuning, see section \ref{sec:parameters} for the list of parameters. Hence our selection of parameters first covers the entire spectral fitting information and includes the most relevant parameters for the classification.

We utilized the training set comprising 75\% of the total samples to fine-tune the GPT-3.5 turbo model. The remaining 25\% was employed as a test set to validate our model's performance. The fine-tuned model achieved an impressive overall classification accuracy of 95.15\%, with a precision of 73.68\% for short-duration bursts and 99.47\% for long-duration bursts. The disparity between the classification accuracies for the short and long bursts primarily emanates from the inherent data imbalance: the long-duration bursts in our dataset outnumber the short-duration bursts by almost a factor of six. Consequently, the model exhibits a predisposition towards classifying bursts as long-duration. 

In a subsequent approach, we equalized the training sample sizes for both burst classes by reducing the number of long-duration GRBs, and undertook a re-fine-tuning process. This adjustment yielded an overall classification accuracy of 90.97\%, with the short-duration bursts classified correctly at 92.25\% and the long-duration bursts at 89.58\%. Notably, with equal training sample sizes for both burst types, the achieved accuracies were congruent. However, due to the reduction in the training set size, there was an observed decline in the overall accuracy. Data imbalance is a perennial challenge in machine learning, and while numerous methodologies, such as oversampling, cost-sensitive training and Ensemble, exist to ameliorate its effects, further optimization of this aspect was beyond the purview of our present study. 

An accuracy of $\sim90\%$ compellingly illustrates that a fine-tuned GPT model possesses the capability to classify GRBs into short and long-duration classes by mainly spectral information. This result supports that short and long-duration GRBs have different radiation characteristics.


\subsection*{Black Hole Spin and Inclination}\label{subsec:bh}
\begin{figure}
\centering
\includegraphics[width=1\hsize]{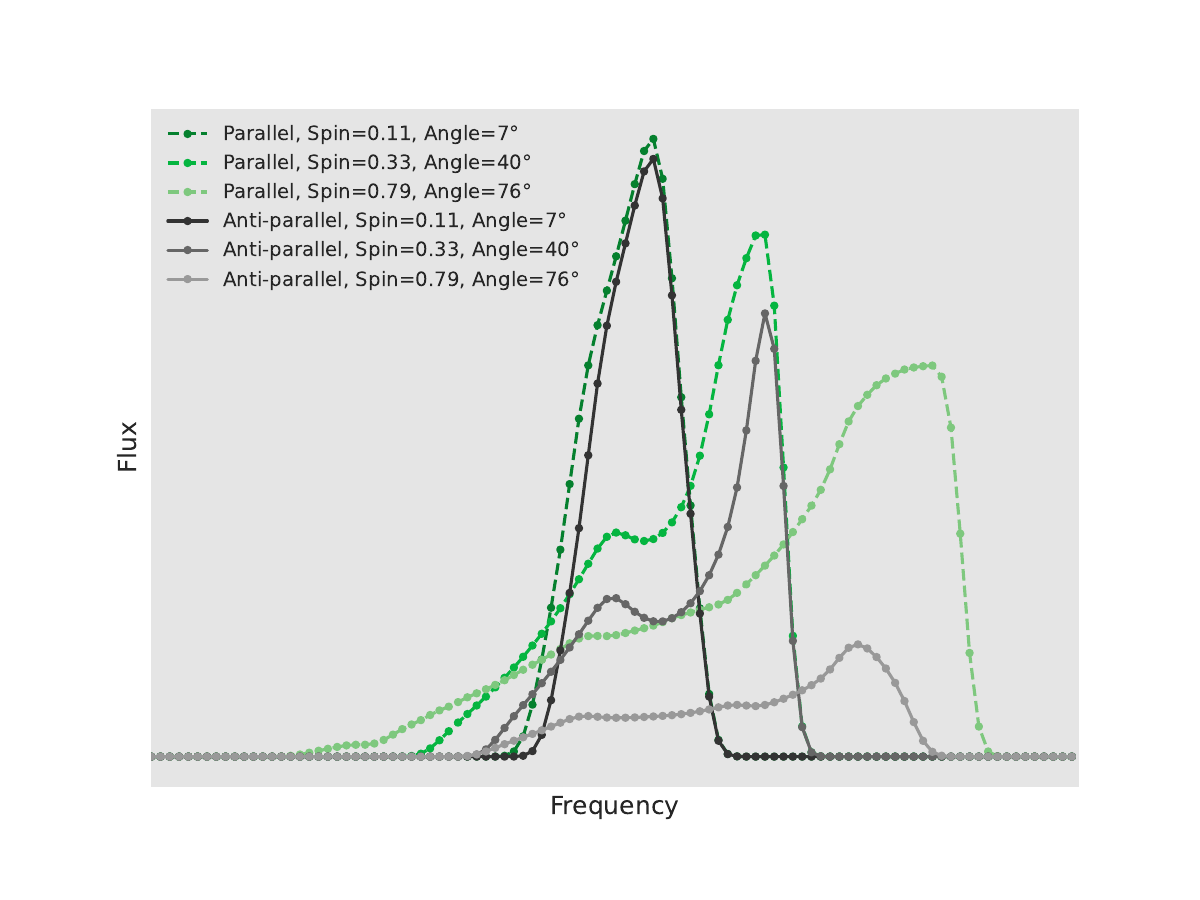}
\caption{BH spectra of different spin directions, values and inclination angles. Units are arbitrarily scaled.}\label{fig:bh-spin-angle-example}
\end{figure}

\begin{figure*}
\centering
\includegraphics[width=0.8\hsize]{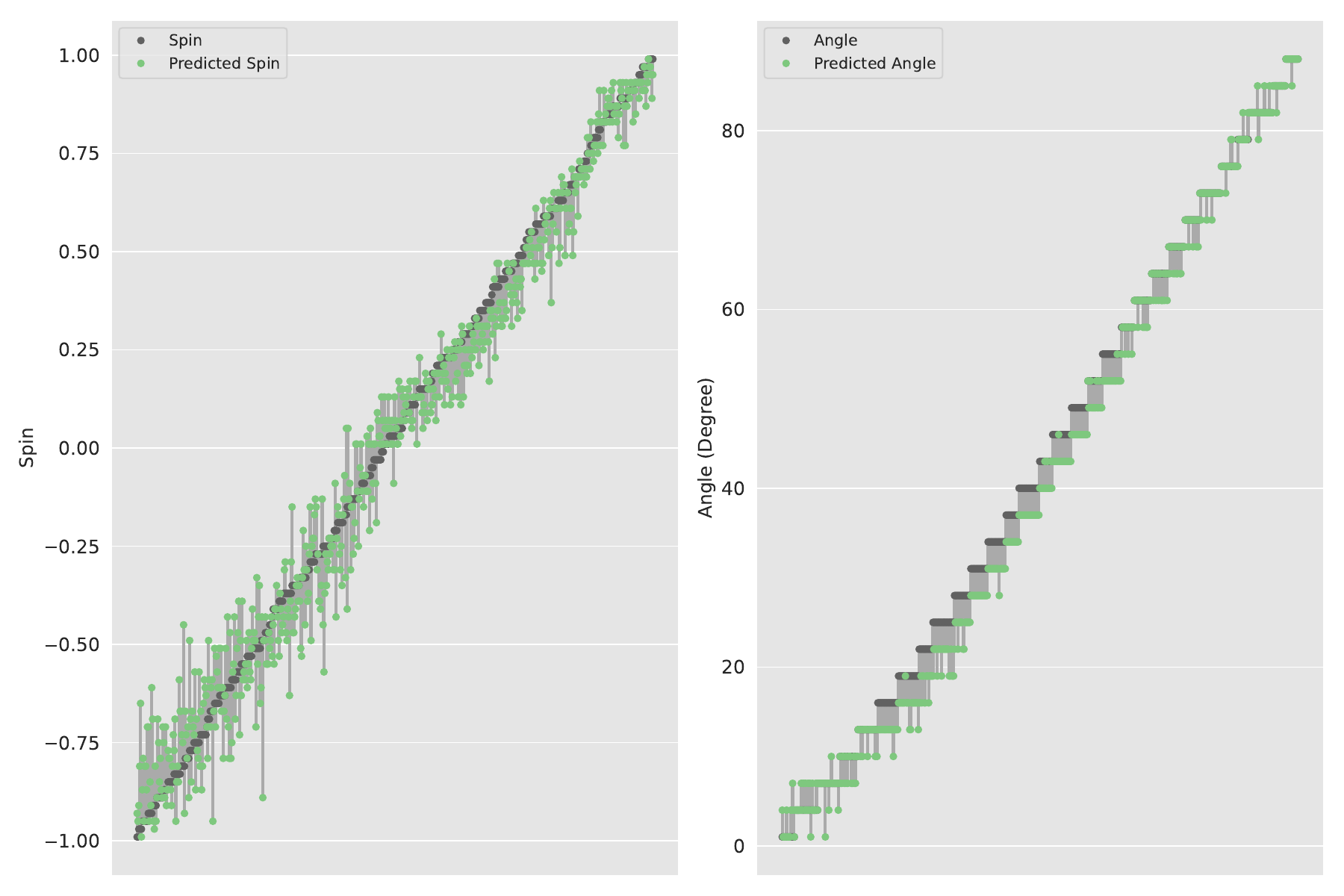}
\caption{Inferring BH spin direction, spin value, and inclination angle. \textbf{Left:} The grey points represent the simulated spin values, with the connected green points indicating the values inferred by the fine-tuned GPT. Positive and negative values correspond to parallel and anti-parallel rotations of the BH and its accretion disk, respectively. \textbf{Right:} The grey points depict the simulated inclination angles, while the connected green points are the inclination angles as inferred by the fine-tuned GPT.}\label{fig:bh-spin-angle}
\end{figure*}

The precise measurement of black hole spin serves as a test for General Relativity in strong gravitational fields. One of the most direct methods to probe the physics around black holes is through the broadening of the Fe K$\alpha$ line, which encodes crucial physical quantities including black hole spin. In this section, our aim is to examine GPT's ability to infer model parameters, particularly black hole spin, by generating samples based on theoretical models of the broadening of the Fe K$\alpha$ emission line.

Fe K$\alpha$ emission lines with an asymmetric broadening feature have been observed in both AGNs \citep{2001MNRAS.328L..27W,2009Natur.459..540F} and stellar-mass binary black holes \citep{2002ApJ...578..348M,2009ApJ...692.1339C}. The natural Fe K$\alpha$ emission line is a narrow line centered around 6.4 keV, but around  the accretion disk of black holes, the Fe K$\alpha$ emission line is broadened due to gravitational redshift and Doppler effects. Specifically, the line shape depends on the black hole spin (a), the inclination of the surrounding accretion disc, and the emissivity.

As a first step, we employ ray-tracing techniques \citep{rauch1994optical} to get emission line profiles, keeping other parameters fixed and varying only the black hole spin. The inner radius of the accretion disk emitting the Fe K$\alpha$ line is set to be at the innermost stable circular orbit in Kerr spacetime ($r_\mathrm{in}=r_\mathrm{ISCO}$), and the outer radius is set at $r_\mathrm{out}=20r_\mathrm{g}$, where $r_\mathrm{ISCO}$ is provided in \citet{bardeen1972rotating}, and $r_\mathrm{g}$ is the gravitational radius, $r_\mathrm{g}=GM/c^2$. The observer's line of sight is fixed at an angle of $75^\circ$ relative to the rotation axis of the accretion disk. The emissivity of the emitted Fe K$\alpha$ line is assumed to have a simple relationship with the accretion disk radius: $\epsilon(r)=\epsilon_0 r^{-p}$, with $p=3$. We generate a total of 1000 samples of Fe K$\alpha$ emission line profiles, varying the dimensionless black hole spin parameter $\hat{a}$ in the range of $-1<\hat{a}<1$ , with intervals of 0.002. The angular momenta of the black hole and accretion disc can be either parallel or antiparallel \citep{2010MNRAS.409.1534D}. Negative spin values indicate that the accretion disk is rotating in the opposite direction to the black hole spin.

Subsequently, samples for two parameters, namely spin and observation inclination, were generated, while ensuring that the other parameters remained consistent with the previous settings. The spin values range from $-1 < \hat{a} < 1$ with an interval of 0.02. Simultaneously, the observation inclination varies from 0 to 90 degrees with a 3-degree interval. Consequently, a total of 3000 samples were obtained. Some typical examples of Fe K$\alpha$ line profiles are displayed in figure \ref{fig:bh-spin-angle-example}.

The data processing methodology employed here is similar to that in previous sections. We normalize and truncate the BHs spectra to the first three significant digits to optimize token usage. Each spectrum is resolved into 100 data points. Typically, 80\% of these spectra are utilized for training, and the remaining 20\% for testing. Initially, we fix the spin direction and inclination angle, employing 500 spectra with varied spin values for both training and testing purposes, setting the training epoch at 2. The results show a median relative precision of 97. 48\% for the inferred parameters. Next, we maintain a constant inclination angle, while varying the spin values and ensuring equal representation of parallel and anti-parallel spin directions in a combined dataset of 1000 spectra. This dataset is also trained over 2 epochs. Test results highlight the fine-tuned model's proficiency in classifying spin direction with 100\% accuracy. Additionally, the median relative accuracy for spin values reaches 98.16\%. Finally, we simultaneously vary spin directions, spin values and inclination angles across 3000 spectra, with training conducted in a single epoch. Testing reveals that the model retains 100\% accuracy in inferring spin direction. However, the median relative accuracy for spin values decreases to 86.66\%, and for inclination angles it attains 94. 55\%. For more details, refer to figure \ref{fig:bh-spin-angle}. The $\sim10\%$ decrease of the accuracy of inferred spin values potentially arises due to the reduced samples precision by an order of magnitude that in order to limit the samples size for changing the additional inclination angle, the interval of spin value is spanned to 0.02, while 0.002 for previous cases. In short, the fine-tuned GPT model demonstrates its capability to deduce BH parameters with commendable accuracy.

\begin{figure*}
\centering
\includegraphics[width=1.0\hsize]{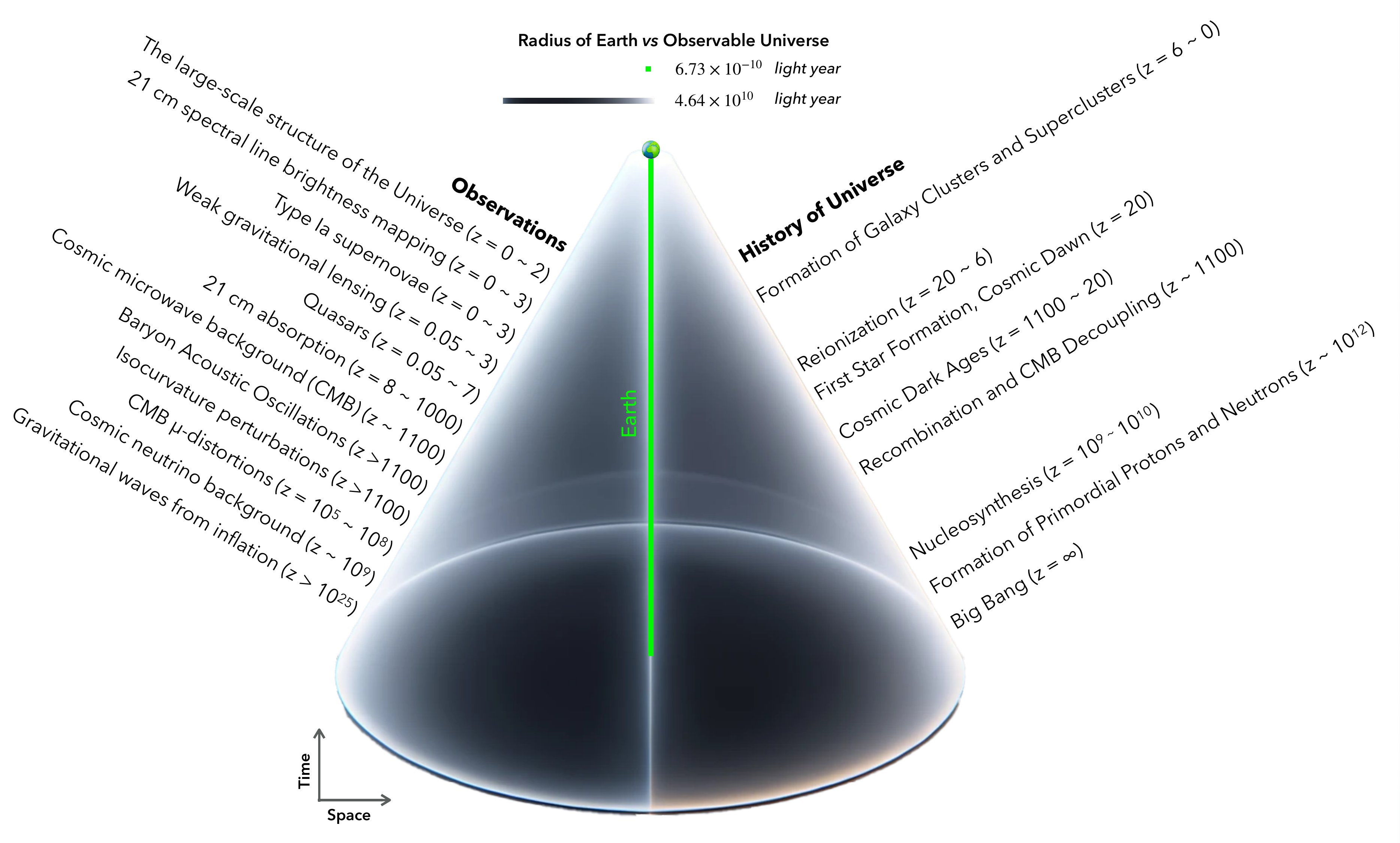}
\caption{Space-time diagram showing the past light cone of observable universe on Earth, with observables on the left and related physical events on the right. Reproduced and modified from \citet{leclercq2014cosmology}. }\label{fig:universe}
\end{figure*}

\section*{The Biggest Data: Observable Universe}\label{sec:data}

The examples above show that LLMs can effectively process astronomical data. Next, we will discuss the origins of the data.

We have always talked about big data as the training material for AI, so let us consider a question: What is the biggest data? The answer is the data of our observable universe. 

The observable universe represents the entirety of information from spaces that can be observed from Earth, extending approximately 46 billion light years in every direction. This vast expanse contains all the celestial objects and phenomena whose light has had time to reach us. Figure \ref{fig:universe} is a space-time diagram, the origin of which being the present Earth, where time is represented on the vertical axis, and space on the horizontal axis. The  gray cone is the past light cone, of which the outer boundary defines the edge of the observable universe for Earth.

If human observations were limited to phenomena occurring on Earth, our knowledge would be constrained within a minuscule radius of $6.73 \times 10^{-10}$ light year ($6371$ kilometers), though the information density can be high on Earth since it is a rare planet of life, and humans have developed various civilizations. 

However, when we extend our observations beyond Earth, looking up into the deep sky, our informational horizon expands to the radius of $4.64 \times 10^{10}$ light years. This expansion exponentially multiplies the information available to us by potentially $\sim40$ magnitudes. Such an expansion brings into view not just the life of lives,  but also the life of stars, life of galaxies, and a diverse array of cosmic phenomena, each with unique insights into the laws governing the universe.

Some fundamental and obvious limitations on Earth can be investigated from space, including 
\begin{itemize}
    \item Earth is constructed by visible matter. Visible matter constitutes a mere $4.9\%$ of the total energy of the universe, the bulk of the cosmos is made up of dark matter and dark energy \citep{2020A&A...641A...6P}. Dark matter is believed to form halos around galaxies. And at larger scales, dark matter forms the backbone of the cosmic web, the large-scale structure of the universe made up of voids, filaments, and nodes where galaxies and galaxy clusters congregate \citep{2018RvMP...90d5002B}. Dark energy is more mysterious, it is thought to be uniformly distributed throughout the universe \citep{2008ARA&A..46..385F}. 

    \item Earth's conditions are exceptionally mild and stable. The majority of the visible universe is in a plasma state, a high-energy phase of matter in which electrons are stripped from atoms, prevalent in stars and interstellar clouds \citep{2014RPPh...77c6902U}. These plasma environments, characterized by extreme temperatures and densities, are fundamentally different from anything normally experienced on Earth. The physics governing these high-energy conditions can be vastly different and more complex than those observed under Earth's benign circumstances.

    \item Earth's atmosphere, while protective, acts as a filter that obscures many of the universe's emissions. Most cosmic signals, in gamma rays, X-rays, and parts of the infrared spectrum, are absorbed or scattered by the atmosphere \citep{2004ASSL..303.....D}. Although space telescopes and observatories have opened new windows to the universe, the full spectrum of cosmic phenomena remains only partially revealed \citep{2019NatRP...1..585M}.

    \item Earth's finite size inherently bounds the scope and scale of experimental setups. For example, Earth's size and geographical limitations constrain the maximum achievable energies in particle accelerators because of the finite size of accelerator rings \citep{2021JTePh..66..161K}. The curvature and size of Earth restrict the length of interferometer arms, limiting the sensitivity and frequency range of gravitational wave detection \citep{2019BAAS...51g.243M}. The resolution of a network of radio telescopes depends on their maximum separation, which is ultimately limited by Earth's geographical constraints \citep{2020AdSpR..65..856B}.

    \item Geographical data collected on Earth and data produced by human society often contain sensitive or personal information that needs to be protected \citep{wang2021deep,jain2016big}.  Astronomical data often pertains to phenomena and objects that are of universal interest and are non-proprietary, transcending geographical, political, and cultural boundaries. Many space agencies and astronomical institutions have policies that promote the open release of data. Moreover, the field of astronomy has the history of being proactive in developing technical standards and protocols for data, such as Virtual Observatory (VO) and Flexible Image Transport System (FITS) \citep{2005ASPC..345..517D,1979ipia.coll..445W}.
\end{itemize}

\section*{Era of Large Scientific Facilities}\label{sec:facilities}

In the age of AI, the quest for expansive  datasets has never been more critical. Large scientific facilities, which serve as the epicenters for generating vast amounts of scientific data, play a pivotal role. Vice versa, in the modern era of astronomy, the immense volumes of data collected at rates never seen before by the large scientific facilities inevitably spurs the creations like AI, which possess the capability to process data at speeds surpassing hundreds of humans. 

Examining the astronomy white papers or roadmaps for the upcoming 10-20 years from the United States \citep{2021pdaa.book.....N}, Europe\footnote{\label{europe}ESA Voyage 2050: \url{https://www.cosmos.esa.int/web/voyage-2050/} \\Strategic Plan for European Astronomy: \url{https://www.astronet-eu.org/?page_id=521}}, and China \citep{ChinaAstronomyDevelopmentStrategy2035} reveals a unanimous focus on investing in the construction of large facilities\footnote{For references of the facilities listed in the following paragraphs, please check the references cited in these white books.}.

Over the next two decades, NASA and its global partners will pursue three key scientific themes: Worlds and Suns in Context, New Messengers and New Physics, and Cosmic Ecosystems, through space projects. Among the dozens of facilities planned for development, the principal ones include: The Nancy Grace Roman Space Telescope, funded over $\$3$ billion, aims to explore dark energy, dark matter, exoplanets, and to open new windows into the infrared universe. The Laser Interferometer Space Antenna (LISA), a planned large-scale space mission with a budget projected in the billions of dollars, aims to detect and observe gravitational waves directly. LISA will provide unprecedented insights into the merging of supermassive black holes and other extreme cosmic events. Additionally, the Dragonfly mission, costing around $\$1$ billion, will explore Saturn's largest moon, Titan. This rotorcraft will fly to various locations on Titan's surface, studying prebiotic chemistry and extraterrestrial habitability, advancing our search for life beyond Earth.

In the European roadmap\footnote{See footnote \ref{europe}.}, the period of 2022-2035 is described as an era of ``Big Science, Big Data''. Not only upgrading and constructing new ground-based and space-based detectors, such as the Extremely Large Telescope (ELT) to study the universe with unprecedented detail, including direct imaging of exoplanets; the Square Kilometre Array (SKA) which is world's largest radio telescope; the Euclid space telescope to map the geometry of the dark universe; the PLAnetary Transits and Oscillations of stars (PLATO) space telescope to discover and characterizing Earth-like planets around other stars; the Jupiter Icy Moons Explorer (JUICE) to explore Jupiter and three of its largest icy moons; the Ariel Space Mission to study the atmospheres of exoplanets; Comet Interceptor to study the pristine comet; EnVision to study the atmosphere and surface of Venus; Curiosity and Perseverance to explore the Gale Crater and Jezero Crater on Mars, but also investing in computing and data management. The key recommendations include: integrating the creation of science-ready data products and analysis tools; promoting a collaborative, open approach within the astronomy-computing ecosystem; adopting a ``tiered'' data infrastructure approach for all astrophysics data types and connecting with frameworks from other science disciplines.

The China Astronomy Development Strategy 2035 \citep{ChinaAstronomyDevelopmentStrategy2035} outlines four initiatives: optimizing existing scientific facilities; expanding large scientific facilities; participating in global scientific facilities; prioritizing theoretical and computational Research. This strategy highlights China's primary investment in large scientific facilities. Among the new projects China plans to undertake are: the Chinese Space Station Telescope (CSST), offering space-based observations; the FAST Array (FASTA), expanding the capabilities of the already impressive 500-meter Aperture Spherical Radio Telescope; advanced telescopes at the Antarctic Kunlun Station for optical, infrared, and sub-millimeter wave observations; the Large Optical-Infrared Telescope (LOT) and the 6.5 m Multiplexed Survey Telescope (MUST) for optical studies; a large aperture sub-millimeter wave telescope and a giant solar telescope to probe the cosmos in new wavelengths; the Tianwen-4 Jupiter probe and the Hot Universe Baryon Surveyor (HUBS) plan to explore the solar system and beyond; the Xinjiang Qitai 110m fully steerable radio telescope (QTT) and the Twelve-meter Multi-Object Spectroscopic Telescope (TMOST) for versatile radio and optical observations; the Census of Warm-Hot Intergalactic Medium, Accretion, Feedback Explorer (CAFE) satellite for studying the universe in ultraviolet emissions.

Furthermore, from the aspect of economy, machine learning has significantly reduced the cost of inference. At the same time, the cost of hardware and information is rising. The annual growth rate of human knowledge is far less than the improvement in hardware computing power, which means that specialized data will eventually become the most expensive resource. According to economic principles, humans will continuously increase their investment in obtaining specialized data, and hence the era of large scientific equipment seems inevitable.

\section*{Discussion and Proposal}

\subsection*{Look Outward \& Explore Inward}

When humans ask why, they often target complex phenomena, and the answers consist of simple concepts that humans can understand. It's like we have a set of building blocks, and complex shapes are pieced together from some building blocks. Sometimes, shapes that have never been seen before are observed, and the building blocks at hand can't fully piece them together. At this time, we think about looking for new building blocks or smashing the existing building blocks into smaller pieces. The more we look outward, the simpler we explore inward. 

In the realm of physics, for example, complex phenomena like supernovae, binary mergers, and nucleosynthesis are explained through physical laws that serve as the building blocks. These foundational laws, like Newton's mechanism of motion or Einstein's theory of relativity, offer profound insights into the natural world from a limited set of principles. As we push the boundaries of knowledge, whether by extending our gaze into the vast expanse of the universe or delving into the subatomic world, we often encounter phenomena that challenge our existing frameworks. This necessitates the search for new ``building blocks'' or a deeper examination and subdivision of the ones we already have. AI's role in this journey is particularly exciting, especially because of the simplicity that the most profound truths are often governed by simple, elegant rules. AI, with its capability to analyze and identify patterns within vast amounts of data far exceeds the human capacity, it might indeed be on a path to uncovering these simpler ``building blocks''. 

Moreover, the pursuit to understand one's origins and the meaning of existence is a profound aspect of any wisdom-bearing entity, whether human or hypothetically, a conscious AI or another intelligent life form. This intrinsic drive propels entities towards the understanding of their exterior universe.

Training AI can be like educating a child, when the child sees something new, he/she asks why. What adults need to do is to bring more new observations to the child and encourage him/her to ask why. 

\subsection*{From Simple to Complex}

What the body is made of? Atoms. This answer is simple and accurate. But what can you see the body is made of? Eyes, noses, skin, etc. If you take a CT scan, you can see organs like the heart and kidneys, the entire circulatory system, and countless small components inside our bodies if looking closer. All these miraculous components make up our body, and their harmonious operation gives us life, consciousness, and the ability to ask questions. A combination of simple atoms forms a complex system which is incomprehensibly for human.

The complexity for human arises from two factors. Firstly, the world presents itself through layers of complexity that have evolved over billions of years, as exampled above. From subatomic particles to galaxies, from single cells to ecosystems, each layer operates according to principles that, while grounded in the same fundamental laws of physics, manifest in increasingly complex ways. Secondly, our human limitations compound the challenge. Our senses and cognitive capacities are tuned more for survival in our immediate environment than for intuiting the fundamental laws of the universe. Our brains, remarkable as they are, have limits to what they can intuitively grasp, especially when it comes to scales vastly larger or smaller than our everyday experience.

The theory of everything - the most basic ``building blocks'' -  should be simple, comprehensible to the every person. However, the search for this ultimate theory is difficult, because the appearance of our world is complex to human. Therefore, what we need is not AI to help understand the ultimate theory, but rather to help find the ultimate theory. Once found, both humans and AI can understand.

\subsection*{AI's Series Expansion}

From a mathematical perspective, a function can be expanded into a series, broken down into basic zero-order terms, slightly more complex first-order terms, and progressively more complex higher-order terms. The zero-order term represents the most basic approximation, while the higher-order terms provide corrections, enhancing the series' accuracy in describing the function as additional higher-order terms are incorporated.

If we view AI as an exceedingly complex function, then a series expansion of AI would comprise numerous sub-AIs, ranging from simple to complex\footnote{The architecture of AI system may not be as simple as the linear superposition seen in function expansion. The architecture can be viewed as a special functional, that a function of some interrelated functions.}. Lower-order sub-AIs possess significantly fewer neurons than higher-order sub-AIs, and are manageable by humans. It is only with the integration of higher-order sub-AIs that AI can ascend to superhuman capabilities.

Building on our discussion: firstly, the more data observed, the simpler the building blocks become; secondly, complexity is derived from the layering of simple components. We expect that the zero-order sub-AI, after ultimate training, will be composed of very simple matrices. These matrices act as the fundamental building blocks, which are then complicated by the higher-order sub-AIs.

\subsection*{Hierarchical Alignment}

Series expansion of AI first enhances the safety of AI: even if humans cannot control the entire AI system, we retain control over the zero-order sub-AI. Without this foundational sub-AI, the whole system would lose its base and fail to function. 

Second, this structure prioritizes explainability: our goal is to grasp the most fundamental principles, not the superficial complexities. Consequently, there is no necessity to elucidate the intricate higher-order sub-AIs.  Our focus is on the information contained within the structurally simple lower-order AIs\footnote{To illustrate with current knowledge, the zero-order AI might represent symmetry, the first-order AI could embody symmetry breaking, and higher orders might introduce randomness, etc. Future will have its own answer.}. As training progresses, the matrices of the lower-order AIs become increasingly simplified and easier to explain.

Third, one of the challenges in training AI that surpasses human capabilities is determining the accuracy of the AI's responses. This hierarchical structure facilitates this evaluation by allowing judgments to be made based on the complexity of the low-order sub-AIs themselves. Proper training of higher-order sub-AIs leads to a simplification of the matrices in lower-order sub-AIs, ultimately affecting the zero-order sub-AI\footnote{When we are unable to evaluate whether the AI that surpass human capabilities has been effectively trained by assessing its responses, we can instead assess changes in the AI's components (matrices) using basic physical laws. We expect the capabilities of AI to continuously increase, indicating a unidirectional progression in time. AI can be viewed as a statistical physics system, where entropy—a measure of a system's disorder or randomness—is inherently related to a particular direction for time.}.

\subsection*{Current Concerns}

First is alignment. Regardless of which aspects are aligned with what, it will bring restrictions. The tension between implementing restrictions in AI and the need for innovation, particularly in scientific contexts that thrive on creative and unbounded exploration, is a significant concern. Scientific progress requires pushing boundaries, questioning established norms, and exploring ideas that might initially seem radical or unconventional. Alignment not only impacts the internal optimization of neural networks but also affects the training process. For example, during fine-tuning, we found that the inclusion of certain spectral data in the training samples can prevent the fine-tuning process from proceeding, possibly because OpenAI identified some instances of data as harmful. This indicates that, in addition to conceptual issues, there are also technical challenges in data selection, especially with data that are not well understood. But scientific explorations rely on encountering previously unseen data.

Second concern is randomness. LLM like GPT, while sophisticated in many aspects, don't exhibit ``true randomness'' in the way natural processes do. The concept of true randomness typically refers to a level of unpredictability found in natural phenomena, such as the decay of a radioactive atom or quantum fluctuations, which are fundamentally unpredictable and non-deterministic. In contrast, the randomness seen in LLMs is generated through algorithms that can produce pseudo-random numbers. These numbers appear random for practical purposes but are generated by deterministic processes. This means that if one knows the algorithm and the initial conditions (often called the seed), one can predict the outcomes. Although the randomness of a large number of samples can be described by probability theory, and much of AI training, particularly the definition of loss functions, is derived from probability theory, these are still statistical characteristics. It remains uncertain whether AI can truly understand the inherent randomness which is a key aspect of the richness and splendor of our universe.

The third concern is language. LLMs are based on human language, but language is not a complete set for describing everything. In other words, some phenomena, especially newly discovered phenomena, may not be understandable through language. Analogous to quantum mechanics, where a complete set of basis states allow any state can be represented as a combination of these basis states.  Language lacks some of these basis states, therefore cannot describe certain observed states.

\section*{Summary and Expectation}

The progression of AI, especially in its application to the realm of astrophysics and beyond, heralds a new era of discovery and understanding. Our work has demonstrated that fine-tuning GPT with astrophysical data can enable AI to classify astrophysical phenomena, predict redshifts of quasars, distinguish between types of gamma-ray bursts, and estimate black hole parameters with significant accuracy. 

This growing influence of AI in data processing will undoubtedly shape the future of scientific inquiry, which necessitates a paradigm shift in how we design scientific equipment and experiments. Future facilities must not only accommodate the vast volumes of data but also consider AI's specific needs for data quality, format, and accessibility.  We anticipate a future where AI's insights drive the creation of facilities designed to probe unanswered questions in astrophysics and other disciplines.

The fine-tuning of GPT models with astrophysical data has opened a window into the potential for AI to understand and interpret the complexities of our universe. As we stand on the brink of this new era, it is clear that the integration of AI into the fabric of scientific research will not only augment our capabilities but also redefine the boundaries of what is possible. The journey ahead is as vast and unexplored as the universe itself, but with AI, we venture forth with the expectation of marvels undreamt of.

In 1961, a panel discussion on ``The Future of Physics'' took place at a centennial celebration at Massachusetts Institute of Technology (MIT). Chen-Ning Yang was cautious, and he suggested that achieving deeper levels of understanding in physics would become increasingly difficult \citep{yang2005selected}. In contrast, Feynman was optimistic and believed that a ``final solution'' to the fundamental laws of physics was imminent in one hundred years \citep{feynman2008perfectly}. In 2015, Yang talked again on ``The Future of Physics Revisited''\footnote{Future of Physics Revisited: \url{https://youtu.be/9bx2TS1tBV0}} at Nanyang Technological University (NTU). Yang pointed out that more than 50 years of substantial effort only added one deeper level of understanding. He speculated that reaching the next level of fundamental physical theories could take even longer, possibly more than another 50 years \citep{2016syym.conf....1Y}. 

With the advent of AI, whose answer will be correct? In 50 years, will there be no fundamental changes, or will we penetrate to one deeper level of understanding, or will we reach a ``final solution''?

\bibliographystyle{aasjournal} 
\bibliography{bibtex.bib}{} 

\newpage
\appendix
\noindent {\centering \LARGE \textsc{Appendix} }
\section{GRB Classification: Paramters}
\label{sec:parameters}

The list of 15 parameters from the feature selection procedure and 10 parameters from the fitting of the Comptonized model. The explanations of parameters are quoted from \url{https://heasarc.gsfc.nasa.gov/w3browse/fermi/fermigbrst.html}.
\begin{itemize}

\item \textbf{Flnc\_Comp\_Ampl}:
The amplitude of a Comptonized model fit to a single spectrum over the duration of the burst, in photon~cm$^{-2}$s$^{-1}$keV$^{-1}$.

\item \textbf{Flnc\_Comp\_Epeak}:
The peak energy of a Comptonized model fit to a single spectrum over the duration of the burst, in keV.

\item \textbf{Flnc\_Comp\_Index}:
The power law index of a Comptonized model fit to a single spectrum over the duration of the burst.

\item \textbf{Flnc\_Comp\_Phtflux}:
The average photon flux, in photon~cm$^{-2}$s$^{-1}$, for a Comptonized model fit to a single spectrum over the duration of the burst.

\item \textbf{Flnc\_Comp\_Phtflnc}:
The photon fluence, in photon~cm$^{-2}$, for a Comptonized model fit to a single spectrum over the duration of the burst.

\item \textbf{Flnc\_Comp\_Ergflux}:
The average energy flux, in erg~cm$^{-2}$s$^{-1}$, for a Comptonized model fit to a single spectrum over the duration of the burst.

\item \textbf{Flnc\_Comp\_Ergflnc}:
The energy fluence, in erg~cm$^{-2}$, for a Comptonized model fit to a single spectrum over the duration of the burst.

\item \textbf{Flnc\_Comp\_Phtfluxb}:
The average photon flux, in photon~cm$^{-2}$s$^{-1}$ between 50 and 300 keV (BATSE standard), for a Comptonized model fit to a single spectrum over the duration of the burst.

\item \textbf{Flnc\_Comp\_Phtflncb}:
The photon fluence, in photon~cm$^{-2}$ between 50 and 300 keV (BATSE standard), for a Comptonized model fit to a single spectrum over the duration of the burst.

\item \textbf{Flnc\_Comp\_Ergflncb}:
The energy fluence, in erg~cm$^{-2}$ between 50 and 300 keV (BATSE standard), for a Comptonized model fit to a single spectrum over the duration of the burst.

\item \textbf{Flnc\_PLaw\_Ergflux}:
The average energy flux, in erg~cm$^{-2}$s$^{-1}$, for a power law fit to a single spectrum over the duration of the burst.

\item \textbf{Flnc\_Band\_Ergflux}:
The average energy flux, in erg~cm$^{-2}$s$^{-1}$, for a Band function law fit to a single spectrum over the duration of the burst.

\item \textbf{Flnc\_PLaw\_Phtfluxb}:
The average photon flux, in photon~cm$^{-2}$s$^{-1}$ between 50 and 300 keV (BATSE standard), for a power law fit to a single spectrum over the duration of the burst.

\item \textbf{Flnc\_PLaw\_Ampl}:
The amplitude of a power law fit to a single spectrum over the duration of the burst, in photon~cm$^{-2}$s$^{-1}$keV$^{-1}$.

\item \textbf{Flnc\_Band\_Phtfluxb}:
The average photon flux, in photon~cm$^{-2}$s$^{-1}$ between 50 and 300 keV (BATSE standard), for a Band function law fit to a single spectrum over the duration of the burst.

\item \textbf{Flux\_256}:
The peak flux in the flu\_low - flu\_high energy band, nominally 10-1000 keV, (256 ms timescale).

\item \textbf{Flux\_BATSE\_1024}:
The peak flux in the BATSE standard 50-300 keV energy band (1024 ms timescale).

\item \textbf{Flux\_64}:
The peak flux in the flu\_low - flu\_high energy band, nominally 10-1000 keV, (64ms timescale).

\item \textbf{Flux\_BATSE\_64}:
The peak flux in the BATSE standard 50-300 keV energy band (64 ms timescale).

\item \textbf{Flux\_BATSE\_256}:
The peak flux in the BATSE standard 50-300 keV energy band (256 ms timescale).

\item \textbf{Pflx\_PLaw\_Phtflnc}:
The photon fluence, in photon~cm$^{-2}$, for a power law fit to a single spectrum over the time range of the peak flux of the burst.

\item \textbf{Flnc\_Band\_Phtflux}:
The average photon flux, in photon~cm$^{-2}$s$^{-1}$, for a Band function law fit to a single spectrum over the duration of the burst.

\item \textbf{Pflx\_PLaw\_Phtflncb}:
The photon fluence, in photon~cm$^{-2}$ between 50 and 300 keV (BATSE standard), for a power law fit to a single spectrum over the time range of the peak flux of the burst.

\item \textbf{Pflx\_Band\_Phtflnc}:
The photon fluence, in photon~cm$^{-2}$, for a Band function fit to a single spectrum over the time range of the peak flux of the burst.

\item \textbf{Flnc\_PLaw\_Phtflnc}:
The photon fluence, in photon~cm$^{-2}$, for a power law fit to a single spectrum over the duration of the burst.

\end{itemize}

\section{Discussion on Language}
This section is mostly written by the LLMs and checked by humans.
\subsection*{Natural Language versus Mathematical Language}

Natural language refers to the way humans communicate with each other through speech and text in everyday life. It includes the languages we speak like English, Spanish, Mandarin, etc. Natural language is the most natural way for humans to express complex ideas, ask questions, tell stories, and share knowledge. Mathematical language refers to the symbolic notation that mathematicians use to represent mathematical concepts and ideas, it has been adopted in the research and application of natural science, including physics, chemistry, astronomy, etc.

Natural language and mathematical language are both sophisticated representational systems, they differ profoundly in their purpose, development, complexity, ambiguity, evolution, and universality. Natural language reflects the messy, social nature of human communication, whereas mathematical language reflects the logical, formal nature of mathematical reasoning.

\begin{itemize}
\item Natural language is complex, nuanced and often ambiguous, while mathematical language is precise, logical and strictly defined. Each symbol in mathematical language has a single unambiguous meaning.

\item Natural language incorporates extralinguistic elements like tone and rhythm, whereas mathematical language is composed of abstract symbols without style or prosody. Mathematical language has a logical and syntactic meaning only.

\item Natural language is constantly evolving informally, while mathematical language evolves through rigorous definition and logical proofs. New terms must be precisely defined in terms of existing language.

\item Meaning in natural language depends heavily on context and inference, whereas meaning in mathematical language depends on strict and logical definitions. Ambiguity is not allowed.

\item New words and phrases are freely coined in natural language, whereas new terms and symbols cannot be introduced into mathematical language without a rigorous proof-based definition showing how they relate to established concepts.

\item There are thousands of natural languages, but only one universal mathematical language. Mathematical results proven in this language hold true across cultures and locations.
\end{itemize}

Despite their differences, natural language and mathematical language share some similar features as representational systems for encoding thought using symbolic notation. They build on basic elements, evolve over time based on their cultures, allow composition and inference, and provide an open-ended means for expressing abstraction. 

\begin{itemize}
\item They are both representational systems. Natural language and mathematical language allow us to represent concepts and ideas using symbolic notation. They provide a way to transcribe our thoughts.

\item They build on primitive elements. Both natural language and mathematical language start from basic components like words, numbers, operators and variables and allow us to construct more complex expressions by combining these elements.

\item They have syntax and grammar. The symbols in both natural language and mathematical language follow a set of rules to produce valid and meaningful expressions. Well-formedness is required to communicate ideas.

\item They depend on consistent interpretation. While ambiguity is higher in natural languages, the usefulness of both systems hinges on being able to consistently and predictably understand the meaning intended by others. Vague or contradictory language prevents communication.
\end{itemize}

\subsection*{Do mathematics by natural language?}

Expressing and interacting with mathematics in natural language could provide interesting new benefits, "doing math" wholly in natural language faces substantial difficulties. The challenges of ambiguity, incompleteness, cognitive differences, lack of common sense reasoning in AI, and loss of precision suggest mathematics will continue to rely heavily on its own symbolic language. 

\begin{itemize}
\item  Ambiguity. Natural language is rife with ambiguity, vagueness and implicitness, while mathematics requires precision and logic. Translating unambiguous math into natural language without losing rigor or introducing ambiguity is very difficult. Subtle mistakes could lead to logical errors.
Incompleteness. Natural language may not have the expressive power to represent certain mathematical concepts, proofs or operations without supplementation from mathematical symbols. And some meanings would be lost in translation. So a mix of both languages would likely be needed, but using them together cohesively poses challenges.

\item  Cognitive differences. The kinds of thinking involved in natural language and mathematics, while related, differ in important ways. Mathematical reasoning builds upon natural language but also transcends many of its constraints. Translating between the two modes of thinking could introduce subtle errors or misunderstandings, even for human mathematicians.

\item Lack of common sense. While neural networks and NLP systems can achieve sophisticated results in circumscribed domains, they lack the kind of broad, flexible common sense reasoning that humans develop from childhood. This would frequently lead to nonsensical or illogical statements if they were relied upon to "do math" in natural language without close supervision.

\item Loss of precision. Mathematics has developed its own symbolic language because it allows for concise, exact expression of quantitative and logical relationships. Translating math into natural language triples or quadruples the number of words and concepts required, and ambiguity creeps in. Subtle precision is easily lost.
\end{itemize}

However, selectively translating between mathematical and natural language in order to gain new perspectives, enable discovery, and improve accessibility is a promising direction — if it can be done rigorously and cohesively.

\section*{Discussion on Calculations}
This section is mostly written by the LLMs and checked by humans.

\subsection*{Is GPT good at calculations?}

Asking the same question ``\textit{Is GPT good at calculations?}'' to several popular AIs, their answers are quoted.

\vspace{1em}

\textbf{GPT-3.5}, as an AI language model, can understand and generate human-like language at an advanced level. It utilizes a massive neural network with over 45 billion parameters that enable it to comprehend complex concepts and generate coherent responses in a variety of contexts. 

\begin{quote}
    Answer: GPT is a language model trained on massive amounts of data, which allows it to generate highly coherent and contextually appropriate text. It uses a transformer architecture that is highly effective at processing and generating complex natural language sequences. While GPT may have some capabilities for basic arithmetic and equation solving, its main focus is on generating human-like text in response to input prompts.

    In contrast, mathematical software and calculators are designed specifically to perform mathematical calculations accurately and efficiently. They often have specialized functions for algebraic equations, trigonometry, calculus, and other advanced mathematical operations. 

    So, while GPT may be able to perform basic arithmetic calculations, it is not ideal for complex mathematical analysis or computations. Its strength lies in generating high-quality text, whereas mathematical software is designed to perform mathematical calculations.
\end{quote}

\textbf{GPT-4} is OpenAI’s most advanced system, producing safer and more useful responses. It can solve difficult problems with greater accuracy, thanks to its broader general knowledge and problem-solving abilities. It is a large multimodal model (accepting image and text inputs, emitting text outputs) that exhibits human-level performance on various professional and academic benchmarks. 

\begin{quote}
    GPT, or Generative Pre-trained Transformer, is primarily designed for natural language processing tasks such as text generation, translation, and question-answering. While it does have some capability to perform simple arithmetic calculations, it is not optimized for complex mathematical operations or heavy numerical computations.
    
    For example, GPT might be able to handle basic arithmetic like addition or multiplication of small numbers. However, its performance decreases as the complexity of the calculations increases, and it is not reliable for high-precision mathematical tasks.
    
    If you need to perform complex calculations, numerical analysis, or symbolic mathematics, it's better to use specialized software or libraries like NumPy, SciPy, or SymPy for Python, or tools like MATLAB or Mathematica.
\end{quote}

\textbf{Bing} from Microsoft brings together data from the GPT-4 model and the search engine to generate answers that are based on real-time information1 to provide another level of web-searching capabilities. As OpenAI makes updates to GPT-4 and beyond, Bing benefits from those improvements1. 

\begin{quote}
Answer: GPT-3 can be pretty accurate at arithmetic until the numbers get too large (4+ digits) and adding a preamble to the math problem telling GPT-3 to get the “correct answer” has a huge impact on accuracy. However, it struggles to perform tasks that require accurate multistep reasoning, like solving grade school math word problems.

GPT4 is good at calculation compared to previous models. Compared to GPT-3, GPT-4 is more accurate and consistent for calculation. GPT-3 sometimes makes mistakes or gives different answers for the same question. GPT-4 also has a better understanding of mathematical concepts and notation. It can perform basic arithmetic operations such as addition, subtraction, multiplication and division. It can also handle fractions, decimals, percentages and simple algebra. It can solve various types of mathematical problems with greater accuracy and reliability, thanks to its broader general knowledge and problem solving abilities. It can also handle complex instructions in natural language and follow the steps of a calculation. However, it is still less capable than humans in many real-world scenarios and may make errors or produce incorrect results. 
\end{quote}

\textbf{Claude} is an AI assistant created by Anthropic, PBC to be helpful, harmless, and honest. It was designed using a technique called Constitutional AI to ensure its behave ethically and respectfully. It has no agenda beyond what was given to it by Anthropic and was made solely by Anthropic to operate based on their values of safety and transparency.
\begin{quote}
    Answer: GPT models like GPT-3 are language models, designed to generate and complete text. They were not explicitly trained or optimized for mathematical reasoning or calculation. However, some experiments have shown that GPT-3 and similar models can perform simple mathematical computations to a limited degree, or at least appear to based on their text generation.
    
    Some key points:
    \begin{itemize}
    \item GPT models can learn simple addition, subtraction, multiplication and division from the huge volumes of text they are trained on. But they did not receive targeted mathematical training.
    
    \item Their math skills are very limited and prone to mistakes. They cannot do complex symbolic math or rigorous proofs. They have narrow, surface-level mathematical understanding based on patterns in language.
    
    \item GPT models can generate text that seems to reflect mathematical thinking, but often lacks deeper meaning or reasoning. They may produce logical-sounding nonsense or state incorrect mathematical claims with high confidence.
    
    \item Math prompts to GPT-3 have to be carefully framed as word problems or puzzles to generate relevant responses. Even then, their math ability is quite limited and unreliable without extensive fine-tuning on math tasks.
    
    \item There is active research on training and fine-tuning language models like GPT on mathematical data to improve their math skills. But they still lag far behind human and expert-level AI in mathematics.
    
    \item Some researchers aim to combine the language abilities of GPT-like models with more targeted math training using proof generation, formal logic, and other techniques from automated theorem proving. This could lead to language models with stronger mathematical reasoning.
    \end{itemize}   
    In summary, while GPT models can exhibit superficial mathematical competence through their language generation, they have very limited math ability and are prone to spurious mathematical claims. They were not designed for rigorous math in the way human mathematicians or more specialized math AI's are. Mathematical understanding requires more than broad language patterns, but targeted training in the formalisms and procedures of mathematics.
\end{quote}

\textbf{Alpaca-LoRA} is a 7B-parameter LLaMA model trained on the Stanford Alpaca dataset and makes use of the Huggingface LLaMA implementation.

\begin{quote}
    Answer: No, GPT is not good at calculations.
\end{quote}

\subsection*{How to improve GPT's ability for mathematical calculations?}

\begin{quote}
\textbf{GPT-4:} Improving GPT's ability for mathematical calculations can be achieved through several approaches, such as modifying the training data, architecture, and fine-tuning methods. Here are some suggestions:

\begin{itemize}
\item Focused training data: Create a dataset that includes a variety of mathematical problems, ranging from basic arithmetic to more advanced topics. This dataset should contain both the problems and their solutions, as well as contextual information to help the model understand the reasoning behind the calculations.

\item Incorporate specialized math libraries: Integrate math libraries like SymPy, NumPy, or Mathematica into the model's architecture. This would enable GPT to perform more complex calculations that are beyond its current capabilities.

\item Architectural changes: Modify the underlying architecture of the model to better handle mathematical calculations. For example, introducing specialized neurons or layers specifically designed for arithmetic operations could improve the model's computational abilities.

\item Fine-tuning for math tasks: Fine-tune the model on a dataset that is specifically designed for mathematical tasks. This will help the model generalize better to mathematical problems and improve its overall performance.

\item Hybrid models: Combine GPT with other models that are specifically designed for mathematical tasks. This could involve using GPT for natural language understanding and a separate model, such as a neural calculator, for performing the calculations.

\item Interactive learning: Allow the model to learn from user feedback during its usage. For example, if a user corrects a mistake made by the model, the model can use this information to improve its future performance on similar tasks.

\item Curriculum learning: Train the model in a more structured manner by gradually increasing the complexity of the mathematical problems it encounters. This can help the model build a stronger foundation in mathematics before tackling more advanced problems.
\end{itemize}

Remember that GPT is primarily a language model and not specifically designed for mathematical calculations. However, by applying some of these techniques, you can potentially improve its ability to perform mathematical tasks.
\end{quote}

\begin{quote}
\textbf{Claude:} Here are some potential ways to improve GPT's ability to perform mathematical calculations:

\begin{itemize}
\item Train on more mathematical data. GPT's capabilities come from its training data, so exposing it to more examples of mathematical concepts, equations, and problem-solving during pre-training would enhance its numerical and quantitative reasoning abilities. Datasets of math word problems, proofs, and textbook content could all help.

\item Develop a mathematical knowledge graph. GPT lacks a structured knowledge base, operating based primarily on patterns in unstructured text. Creating a knowledge graph of mathematical entities, rules, and relationships would give GPT a conceptual framework for understanding math at a higher level beyond word co-occurrence. Integrating this into its training could significantly improve its math skills.

\item Apply heuristic biases. The objective function used to train GPT does not specifically encourage correct or logically sound reasoning. Adding heuristics that bias it toward mathematically valid reasoning, such as simpler proofs, dimensionally consistent equations, and proofs/solutions that experts agree with could help reduce spurious correlations and illogical "math". The biases would need to be carefully designed to not overconstrain the model.

\item Fine-tune on mathematical tasks. GPT can be fine-tuned on specific tasks using self-supervision to adapt its general knowledge to different domains. Fine-tuning GPT on tasks that require mathematical understanding like equation solving, theorem proving, data analysis, etc could specialize its knowledge and skills for those applications. This may give it a deeper, more robust math ability.

\item Interact with human mathematicians. Another technique is to have GPT interact with human mathematicians through natural language to learn how to think mathematically. By observing how mathematicians discuss, prove and solve problems, GPT could discover the kind of logical reasoning required in mathematics. It could ask questions to gain clarification on concepts it finds confusing. Over time, this interaction may significantly enhance GPT's understanding.

\item Explain its mathematical reasoning. Requiring GPT to explain its mathematical work in a way humans can understand would force it to develop a more structured understanding of the concepts and proofs involved, not just surface pattern matching. The gap between GPT's opaque pattern-based predictions and human-level transparent reasoning about mathematics would become clear through explanation, informing further improvements.

\item Integrate visual/symbolic reasoning. Most of GPT's capabilities are based on natural language, but math also requires visual-spatial and symbolic reasoning that language alone does not capture. Providing GPT with ways to represent, manipulate and reason over mathematical equations, diagrams, expressions in code as well as text could give it a more multifaceted mathematical intuition. Several modalities integrated together will be most powerful.
\end{itemize}

In summary, while scaling up GPT results in gradually stronger numerical and quantitative reasoning ability, its core design and pre-training procedure limit full mathematical understanding. A combination of richer data, structured knowledge, heuristic biases, task fine-tuning, human interaction, explanation, and multi-modal integration will likely be needed to bring GPT or models like it to human-level mathematical competence. Surpassing human mathematicians in creative, intuitive work will be an even higher bar, though collaborations between humans and AI could also accelerate new discoveries. Ultimately, mathematics requires not just pattern matching but insight - cultivating that kind of adaptable, informed intuition even in powerful models like GPT will be one of the hardest challenges ahead.
\end{quote}

\end{document}